\documentclass[12pt]{amsart}
\usepackage{amsmath}
\usepackage{amsfonts}
\usepackage{amssymb}
\usepackage{graphicx}

\newcommand{\beq}{\begin{equation}}
\newcommand{\eeq}{\end{equation}}

\newtheorem{theorem}{Theorem}
\newtheorem{lemma}{Lemma}

\newcounter{one}
\newcounter{two}
\newcounter{six}
\begin{document}
\begin{flushright}
Dedicated to Henry McKean
\end{flushright}

\title{Reality problems in the soliton theory}
\author{P.G.Grinevich}
\author{S.P.Novikov}
\address{L.D.Landau Institute for Theoretical Physics}
\email{pgg@landau.ac.ru}
\address{Institute for Physical Science and Technology, University of Maryland
at College park and L.D.Landau Institute for Theoretical Physics}
\email{novikov@ipst.umd.edu}

\begin{abstract} This  is a survey article dedicated mostly to the theory of real regular
``finite-gap''  ( algebro-geometrical)  periodic and quasiperiodic Sine-Gordon solutions.
Long period this theory remained unfinished and ineffective, and by that reason practically had no applications.
Even for such simple physical quantity as ``Topological Charge'' no formulas
existed expressing it through the ``Inverse Spectral Data''. Few years ago the present
authors solved this problem and made this theory effective. This article contains
description of the history and recent achievements.
It describes also the reality problems for several other fundamental soliton systems.
\end{abstract}
\maketitle
\section{Introduction}

The most powerful method for constructing explicit periodic and quasiperiodic solutions
 of soliton equations
is based on the Finite-gap or Algebro-geometrical approach, developed by Novikov (1974),
 Dubrovin, Matveev,
Its, Lax, McKean, Van-Moerbeke (1975) for
$1+1$ systems, extended  by Krichever  (1976) for $2+1$ systems like KP (see \cite{Nov1},
\cite{IM1}, \cite{DMN},
\cite{L}, \cite{McK-VM} for extra information). Already in 1976 new ideas were formulated
 how to extend this approach to the $2+1$ systems associated with spectral theory of the
 2D Schrodinger operator restricted to one energy level
  (see \cite{Man, DKN} ). These ideas were developed  in 1980s
by several people in the Moscow Novikov's Seminar  (see below). The ``Spectral Data''
characterizing the
associated Lax-type operators consist of Riemann Surface (``Spectral Curve'') equipped
by the selected set of points (``Divisor of Poles'', ``Infinities''). In the finite gap
 case this Riemann surface has finite genus, and the number of selected point  is also finite.
  The algebro-geometric approach
in particular  allows to write down explicit solutions
in terms of the Riemann $\theta$-functions.

In modern literature very often the problem is assumed ``more or less'' completely solved
if such formulas are derived. However, in some cases this belief is too naive and does not
correspond to the needs of real life. For example, it is necessary to select physically  or
geometrically relevant classes of solutions corresponding to the source problem
(i.e. solutions, satisfying some reality conditions, regular solutions, bounded solutions
and so on).
Is it easy or not?

 To reach this goal, following problems  should be solved.
\begin{itemize}
\item Problem 1. How to select solutions real for real $(x,t)$?
\item Problem 2. How to select real nonsingular solutions?
\item Problem 3. How to select periodic solutions with given period (or quasiperiodic solutions
with given group of quasiperiods)?

\end{itemize}{\bf Remark.} We call  solution  {\bf non-singular}, if it is
non-singular on the whole real Abel torus.  It should remain non-singular under action of all
(real) higher
flows from the corresponding integrable hierarchy.

The generic $x$-direction is normally ergodic in the Abel
torus, so this definition is equivalent to the standard one. However, for some specific
values of constants of motion theoretically we may have solutions, which are regular
in the standard sense, but blow-up under the action of the higher symmetries.

For some models like Korteveg-de Vries equation (KdV), defocusing Noninear Schr\"odinger Equation
 (NLS), Kadomtsev-Petviashvity~2 (KP2) equation selection of real and nonsingular solution is rather
straightforward.
But for many other models like focusing NLS, Sine-Gordon equation (SG), KP1, inverse scattering
transformation for
2-D Schr\"odinger operator based at one energy the problem of selecting real solutions is
 rather difficult.

The theory of $\theta$-functions is very complicated and ineffective. The complexity is hidden
 behind the
simple notations in these formulas.

Our goal is to discuss in more details the Sine-Gordon equation:
\beq
\label{SG1-ch1}
u_{tt}-u_{xx}+\sin u(x,t)=0.
\eeq
In the light-cone coordinates
\beq
x=2(\xi+\eta), \ \ t=2(\xi-\eta),
\eeq
it has the form.
\beq
\label{SG-0.2-ch1}
u_{\xi\eta}=4\sin u, \ \ \ u=u(\xi,\eta).
\eeq
According to our definition, the solution $u(x,t)$ is $x$-periodic with period $T$
if $\exp\{iu\}$ is $x$-periodic with that period. For the function $u$ we have
$$
u(x+T,t)=u(x,t)+2\pi n,n\in {\mathbb Z}.
$$
We call the quantity $n$ a {\bf Topological Charge} corresponding the the period $T$.

We call the ratio $n/T$ a {\bf Density of Topological Charge}.

The Density  of Topological Charge can be naturally extended
to all real generic regular finite-gap (quasiperiodic) solutions. It is the most basic
conservation law.

{\bf Problem:} How to calculate  topological charge of the real
finite-gap solutions in terms of the spectral data.

Let us remind, that the Inverse Scattering (Spectral) Data for KdV and Sine-Gordon systems
consist of a Riemann surface (Spectral Curve) $\Gamma$ with finite genus equals to
$g$  and a collection of points (``divisor'') $D=\gamma_1+\ldots+\gamma_g$.
(For NLS and some other systems number of poles may be different from genus).

In the case of KdV (or finite-gap periodic Schr\"odinger operator $L=-\partial_x^2+u(x)$)
this surface  $\Gamma$ is
hyperelliptic. In the case of the Sine-Gordon equation the surface is
 also hyperelliptic,
$\mu^2=\lambda\prod\limits_{i=1}^{2g}(\lambda-\lambda_i)$ with branching points
$(0,\lambda_1,\ldots\lambda_{2g},\infty)$. However the classes of admissible Riemann
surfaces and divisors for KdV and Sine-Gordon are dramatically different (see below).

The $\theta$-functional formulas for Sine-Gordon were obtained by Kozel, Kotlyarov and Its
 \cite{KK},
\cite{IK} in 1976. The reality problem remained unsolved. Indeed, the class of admissible Riemann
surfaces was found in these works (see \cite{IK}). The nonzero finite branching points
$(\lambda_1,\ldots\lambda_{2g})$ can be  either real negative
$(\lambda_1,\ldots\lambda_{2k})\in{\mathbb R}$ or  complex conjugate with nonzero imaginary part
$\lambda_{2k+1}=\bar\lambda_{2k+2}$, \ldots, $\lambda_{2g-1}=\bar\lambda_{2g}$.
However, no ideas were proposed where  the poles are located on the Riemann Surface.

As it was realized in early 1980s by McKean \cite{McK}, Dubrovin, Novikov and Natanzon \cite{DN},
\cite{Dubr-Nat}, Ercolani, Forest and McLaughlin \cite{Erc-Forest},\cite{EFM},
this problem is nontrivial. By that reason periodic finite-gap
Sine-Gordon theory long period had no applications.

An important idea how to describe position of poles for the  real nonsingular solutions was
 in fact suggested by Cherednik
in 1980 \cite{Cher}. He was the first author who discovered (ineffectively) that for the given
admissible real Riemann Surface there can be many different real Abel tori generating real
nonsingular quasiperiodic solutions. Their number is equal to $2^k$ where $2k$ is the number
of negative real branching points. All real finite-gap solutions are nonsingular for Sine-Gordon
for the generic Riemann surface. His work was written in the abstract algebro-geometric form,
and he never
developed
these ideas later.  Extending this approach on the basis of ``Algebro-Topological'' ideas,
  Dubrovin and Novikov
\cite{DN} presented an interesting idea how to calculate  topological charge in terms of the
``inverse spectral data''. However,  as it was
pointed out by Novikov in 1984 \cite{Nov}, there was mistake in their arguments: the formula
 proposed in \cite{DN} was proved
only for the small neighborhood of some very special solutions. The problem remained open till
2001. The complete solution (confirming Dubrovin-Novikov formula) was obtained by the authors
 in \cite{GN} as a development of
the ``Algebro-topological approach'', suggested in \cite{DN}, see also \cite{GN2}, \cite{GN3}.
It is interesting that in the works \cite{Dubr-Nat} and later \cite{ Erc-Forest},
these components were described as the real subtori in
the Jacobian  variety $J(\Gamma)$. However this ``$\theta$-functional description''
did not led yet to any formula for the topological charge.  It does not require any
specific basis of cycles. As  we know now, good formula for the topological charge can be
written in very specific basis only.  We believe that using this basis of cycles one can deduce our
formula from the $\theta$-functional expression. It would be good to do that.

\section{Physically relevant classes of  solutions for the different Soliton Systems}

The Korteveg-de Vries equation (KdV)
\beq
\label{Kdv}
u_t+u_{xxx}-6uu_x=0, \ \ u=u(x,t),
\eeq
was originally derived in the water waves theory. As it was discovered in early 1960s (see
introduction to the book  \cite{ZMNP}), it naturally
appears as a first non-vanishing correction for the dispersive nonlinear  systems if
dissipation can be neglected. In these models only real
non-singular solution are physically relevant.

Integration of  KdV equation is based on the ``Inverse Scattering Transform''
 for the 1-dimensional
Schr\"odinger operator
\beq
\label{Schr-ch2}
L=-\partial_x^2 + u(x,t).
\eeq

Selection of real KdV solutions is rather straightforward.
\begin{enumerate}
\item The spectral curve $\Gamma$ $\mu^2=R_{2g+1}(\lambda)$ should be real. It means, that
$R_{2g+1}(\lambda)=\lambda^{2g+1}+\sum\limits_{i=0}^{2g}p_i\lambda^i $ has real coefficients
$p_i\in{\mathbb R}$, or equivalently, all roots are either real or form complex conjugate
pairs. Therefore we have a holomorphic involution $\tau:(\lambda,\mu) \rightarrow (\bar\lambda,-\bar\mu)$
on $\Gamma$.
\item The divisor $D$ should be real with respect to $\tau$: $\tau D=D$, or equivalently,
the unordered set of points $\gamma_1$,\ldots, $\gamma_g$ is invariant with respect to $\tau$.
Of course, $\tau$ may interchange some of them.
\end{enumerate}

Real nonsingular KdV solutions correspond to the following special spectral data:
\begin{enumerate}
\item All branching points
of $\lambda_k$ of $\Gamma$ are real and distinct. Assume, that $\lambda_1<\lambda_2<\ldots<\lambda_{2g+1}$.
Then $\tau$ has exactly $g+1$ real ovals over the intervals $a_0=(-\infty,\lambda_1]$,
$a_1=[\lambda_2,\lambda_3]$, \ldots,$a_g=[\lambda_{2g},\lambda_{2g+1}]$.
\item Each finite oval
$a_k$, $1\le k \le g$ contains exactly one divisor point $\gamma_k\in a_k$.
\end{enumerate}

{\bf Remark.} A real curve of genus $g$ may have at most $g+1$ real oval. Curves with $g+1$ real ovals
(maximal possible number) are called $M$-curves.

Generic finite-gap solutions are quasiperiodic with $g$ incommensurable periods. How to select
$x$-periodic solutions with prescribed period T?
Avoiding any use of Algebraic Geometry and Riemann Surfaces, nice approach to the characterization of
the  strictly $x$-periodic solution in terms of the so-called ``quasimomentum map''
was developed by
Marchenko and Ostrovskii in 1975 \cite{MO}. This map was studied in the Quantum Solid State Physics Literature
in 1959 (see \cite{Kohn}).  It is well-defined in the upper half-plane
outside of some vertical edges.
Its analytical properties  were effectively used in \cite{MO}.  For example  the
 approximation of $x$-periodic solution (potential) by the finite-gap ones periodic with
 the same period, was proved. Another approach based on the so-called
isoperiodic defomations of finite-gap potentials
was developed by Grinevich and Schmidt in 1995 \cite{GS}. In the KdV case the
isoperiodic deformations can be
interpreted as the so-called ``Loewner equations''
 for the corresponding conformal map.
Let us point out, that there exists a big literature, dedicated to the KdV  solutions with real
poles
(rational solutions, singular trigonometric and elliptic solutions)
-- see \cite{AirMM} where these ideas were started.
These solutions are very important from the mathematical point of view:  for example,  the dynamics
of poles satisfies to the equations of  the rational and elliptic
Moser-Calogero models respectively. However, they are related neither to nonlinear wave problems
nor to the spectral theory of the corresponding Schr\"odinger operators. So we do not discuss this
literature in the present survey article.

The modified Korteveg-de Vries equation has the form:
\beq
\label{MKdv-ch2}
v_t+v_{xxx}-6v^2v_x=0, \ \ v=v(x,t).
\eeq
It is connected with KdV by the Miura transformation:
\beq
\label{Miura-ch2}
u(x,t)=v_x(x,t)+v^2(x,t).
\eeq
The real non-singular solutions are physically relevant.

The ``complex'' Non-linear Schr\"odinger equation (NLS) is a system of equations for the pair
of independent complex functions $q=q(x,t)$, $r=r(x,t)$:
\beq
\label{cNLS-ch2}
\left\{
\begin{array}{l}
iq_t+q_{xx}+ 2q^2 r =0 \\
-ir_t+r_{xx} + 2r^2q =0
\end{array}\right.
\eeq
This system has 2 natural real reductions: defocusing NLS: $r(x,y)=-\overline{q(x,y)}$
\beq
\label{dNLS-ch2}
iq_t+q_{xx}- 2|q|^2q = 0,
\eeq
and self-focusing NLS $r(x,y)=\overline{q(x,y)}$
\beq
\label{sNLS-ch2}
iq_t+q_{xx}+ 2|q|^2q = 0.
\eeq
These equation describes nonlinear media with dispersion relations depending on the square
of the wave
amplitude (see \cite{ZMNP}). Among the todays applications of NLS  is the theory of  light
propagation in the fiber optics.
The sign $+$ or $-$ is determined by the dispersion relation, and the
qualitative behaviour
critically depends on it. From the mathematical point of view, the defocusing NLS system
is much simpler
because the linear Lax operator is self-adjoint. The focusing NLS is much more complicated.
In both cases physical applications requires regular solutions.

The complex NLS spectral data are  following: A hyperelliptic Riemann surface $\Gamma$
with $2g+2$ finite
branching points $\lambda_1$,\ldots,$\lambda_{2g+2}$ and $g+1$ divisor points $D=\gamma_1+\ldots+\gamma_{g+1}$.
In contrast with the KdV case, there is no branching at $\infty$.

Solutions of the defocusing NLS correspond  to the following  spectral data:
\begin{enumerate}
\item{ $\Gamma$ } is real, i.e. the polynomial $R_{2g+2}=\prod\limits_{k=1}^{2g+2}
(\lambda-\lambda_k)$
has real coefficients. $\Gamma$ is defined by $\mu^2=R(\lambda)$. The anti-holomorphic
involution on $\Gamma$
is defined by the map  $\tau: {\mathbb C}^2\rightarrow {\mathbb C}^2$ where
$(\lambda,\mu) \rightarrow (\bar\lambda,-\bar\mu)$.
\item The divisor $D$ is real with respect to $\tau$: $\tau D=D$.
\end{enumerate}
Selection of regular solutions is also very similar to the KdV case
\begin{enumerate}
\item All branching points of $\Gamma$ are real. Therefore $\Gamma$ has $g+1$ real ovals over
 the intervals
$[\lambda_{2k-1},\lambda_{2k}]$, $k=1,\ldots,g+1$, i.e. $\Gamma$ is an $M$-curve.
\item There is exactly one divisor point at each real oval.
\end{enumerate}

Selection of $x$-periodic solutions is completely analogous to the KdV case.

Let us describe  the data generating real solutions of the self-focusing NLS equations
for regular spectral
curves. By the Cherednik theorem \cite{Cher}, these solutions  are automatically
non-singular.
Solutions, corresponding to singular spectral curves can be obtained as proper degenerations.
 In contrast
with the defocusing case, singular curves may generate regular $x$-quasiperiodic solutions.

\begin{enumerate}
\item $\Gamma$ is a real hyperelliptic surface of genus $g$ with $2g+2$ $\lambda_1<\lambda_2<\ldots<\lambda_{2g+2}
$ finite branching points. There are no branching points on the real line, so they form complex conjugate pairs.
The antiholomorphic involution $\tau$ acts on the $\lambda$-plane  as $\tau\lambda=\bar\lambda$.
The points in $\Gamma$ lying over the real line  are invariant with respect to $\tau$. Equivalently,
$\tau:(\lambda,\mu) \rightarrow (\bar\lambda,\bar\mu)$.

\item There exists a meromorphic differential $\Omega$ such, that:
\begin{itemize}
\item $\Omega=(1+o(1))d\lambda$ at the infinite points of $\Gamma$.
\item $\Omega$ is regular outside infinity. Therefore it has exactly $2g+2$ zeroes.
\item Let $D=\gamma_1+\ldots+\gamma_{g+1}$. Then the divisor of zeroes of $\Omega$ is
$D+\tau D$. Therefore, $D+\tau D=2\infty_1+2\infty_2-K$.
\end{itemize}
\end{enumerate}

The Sine-Gordon equation in the light-cone variables was derived in the end of the 19-th century.
It describes immersions of the negative curvature surfaces into ${\mathbb R}^3$. Assume, that
an asymptotic coordinate system is chosen (a coordinate system such, that coordinate lines have
zero normal curvature). The  angle between the coordinate lines satisfy (\ref{SG-0.2-ch1}).
It means, that only real regular solutions such that $u(x,t)\ne 0 (\mod \pi)$, are relevant.

The Sine-Gordon equation describes also  dynamics of the Josephson junctions. In this model
$u(x,t)$ is the phase difference between the contacts, therefore the real non-singular solutions are
relevant. However, according to  the leading experts in the Superconductivity Theory, the problem always
requires boundary problem, so we have to consider either the finite interval or the half-line.

The elliptic Sinh-Gordon equation
\beq
\label{Sinh-Gordon}
u_{xx}+u_{yy}+4 H \sinh u =0.
\eeq
describes the constant mean curvature surfaces with genus equal to one, outside umbilic points (see review \cite{Bobenko}).
The constant mean curvature tori has no umbilic points, therefore real nonsigular solutions
should be selected. In contrast
with soliton equations, all real smooth double-periodic solutions are automatically
finite-gap here: \cite{Hitchin} ,\cite{PS}. It follows from the following observations by Hitchin
\cite{Hitchin}: all isospectral flows from the corresponding hierarchy are zero eigenfunctions
of the linearized problem. But the linearized system is the 2-dimensional (elliptic) Schr\"odinger operator,
and it may have only finite-dimensional space of double-periodic zero eigenfunctions. It means,
that the hierarchy contains only finite number of linearly   independent flows at this point.
As a corollary the spectral curve has finite genus. A further development of this idea was used
by Novikov  and Veselov in the paper \cite{NV3}. It was shown, that all periodic chains of
Laplace transformations consisting of  the 2-dimensional double-periodic Schr\"odinger operators
with regular coefficients are algebro-geometric (2D analogs of ``finite-gap'' operators).

The Boussinesq equation
\beq
\label{boussinesq-ch2}
\left\{
\begin{array}{l}
u_t=\eta_x \\
\eta_t = - \frac13u_{xxx} + \frac43uu_x.
\end{array}\right.
\eeq
is used for describing the water waves. For physical applications it is necessary to select
real non-singular solutions. We would like to point out, that the problem of selecting such
solutions in terms of the finite-gap data remains open.

The Kadomtsev--Petviashvili (KP) equation
\beq
\label{KP-ch2}
(u_t+u_{xxx}-6uu_x)_x+3\alpha^2u_{yy}=0 , \ \ u=u(x,y,t), \ \ \alpha^2\in{\mathbb R}.
\eeq
The auxiliary linear operator for KP has the form
\beq
L= \alpha\partial_y - \partial_x^2 +u(x,y,t).
\eeq
If $\alpha$ is  imaginary, we have the so-called KP1 equation, and $L$ is the one-dimensional
non-stationary Schr\"odinger operator. If $\alpha$ is  real, $L$ is the parabolic
operator. In  both cases the real non-singular solutions are physically relevant only. The necessary
and sufficient conditions for the finite-gap spectral data selecting the real non-singular
solutions were found by Dubrovin and Natanzon \cite{Dubr-Nat2}.

Real non-singular solutions of the KP-2 equation correspond to the following geometry:
\begin{enumerate}
\item $\Gamma$ is a algebraic surface of genus $g$ with a marked point and an antiholomorphic involution $\tau$
such, that the marked point is invariant under the action of $\tau$. The marked point is the essential
singularity of the wave function.
\item $\tau$ has exactly $g+1$ fixed oval, i.e. $\Gamma$ in an $M$-curve with respect to $\tau$. Denote
the oval containing the essential singularity by $a_0$ and the other ovals by $a_n$, $n=1,\ldots,g$.
\item Each oval $a_n$, $n\ne 0$ contains exactly one divisor point.
\end{enumerate}

In the case of the Kadomtsev-Petvishvili 1 equation the reality constraints on the spectral curve are
exactly the same as in the  KP 2 case, but the divisor $D$ has a completely different description:
There exists a meromorphic differential $\Omega$ with exactly one second-order pole located at the marked
point such, that the divisor of zeroes of $\Omega$ is exactly $D+\tau D$. Equivalently, $D+\tau D=2\infty -K$,
where $\infty$ denotes the marked point.
Regular real solutions are generated by the data with the following extra constraint:

The pair $(\Gamma,\tau)$ is of separating type, i.e. after removing all real ovals $\Gamma$ splits into
2 components.

An important example of ``solvable'' inverse spectral transform is the one-energy problem for
the two-dimensional Scr\"odinger operator started in the works \cite{Man,DKN}.
\beq
\label{Sch-ch2}
L=-\partial_x^2 -\partial_y^2 +u(x,y),
\eeq
It is well-known, that the full set of the scattering data  for multidimensional
Schr\"odinger operators $n>1$ is overdetermined. A lot of people studied this problem.
We even don't quote this literature.  However, the case $n=2$ turned out to be very specific.
In 1976 Manakov, Dubrovin, Krichever and Novikov \cite{Man,DKN} started completely
new approach for this specific case: They started the Inverse Scattering theory and corresponding Soliton Theory
associated with one selected energy level.  A lot of work was done later in this subject
later (see \cite{NV, VKN,GN4} and review \cite{Grin} for additional references).
In particular, in the first  work \cite{DKN} they
defined the natural analogs of  finite-gap potentials for the 2-dimensional
problem as the  potentials, ``finite-gap at one energy''.  Let $u(x,y)$ be
double-periodic. Denote the dispersion relation by $\epsilon_j(k_x,k_y)$. The Fermi-curve at the
energy level $E_0$ is defined by:
\beq
\label{Sch2-ch2}
\epsilon_j(k_x,k_y)=E_0
\eeq
Denote the complex continuation of the Fermi-curve by $\Gamma$. The Potential $u(x,y)$ is called
{\bf finite-gap at one energy}, if  $\Gamma$ has finite genus.

For generic spectral data  the operators constructed in \cite{DKN} have generically a non-zero magnetic field, i.e.
they have some extra first-order terms:
\beq
\label{Sch3-ch2}
L=-\partial_x^2 -\partial_y^2+A_1(x,y)\partial_x+ A_2(x,y)\partial_y +u(x,y),
\eeq
It might happen that $H(x,y)\ne0$, where $H(x,y)=\partial_x A_2(x,y)-\partial_y A_1(x,y)$.
 For physical applications it is important
to select the  case of ``potential operators'' $A_1(x,y)=A_2(x,y)=0$ with real potential $u(x,y)$.
Sufficient conditions on the spectral data leading to the  potential operators were found
in 1984 by Novikov and Veselov in \cite{NV}. For double periodic
potentials the existence of such form is necessary. It follows from the direct spectral theory,
developed by Krichever \cite{Krich-twod}.
For the generic  regular quasiperiodic  potentials ``finite-gap for one energy level'',  this problem
remains open. Selection of real potentials here
is simple.

How to select the class of regular potentials in terms of algebro-geometrical spectral data?
 There is no complete solution to this Problem.
It was shown in \cite{NV}, that if the spectral curve is the so-called $M$-curve, then the potential
$u(x,y)$ is regular, and the operator $L$ is strictly positive (the selected energy level lies  below the ground state).
An alternative proof of the last statement was obtained by the authors in \cite{GN4}. The complete
characterization of the data generating strictly positive operators (with real regular potentials) was
``more or less'' clarified but some special features remain
 unproved rigorously.

If the selected energy level  is located above the ground state, the topology of the spectral curve $\Gamma$
become more complicated. Many classes of spectral data generating real non-singular solutions were found
by Natanzon (see review \cite{Nat1}), but the classification is not complete till now.

\section{Sine-Gordon equation.}

Connections between the Sine-Gordon equation and the inverse scattering method were first established
by G.Lamb in 1971 \cite{Lamb}. The modern approach developed by Ablowitz, Kaup, Newell and Segur in 1974 \cite{AKNS}
is based on the following zero-curvature representation:
\beq
\Psi_x=\frac14 (U+V)\Psi, \ \ \Psi_t=\frac14(U-V)\Psi,
\label{3.1}
\eeq
where
\beq
U=U(\lambda,x,t)=\left[\begin{array}{cc}
i(u_x+u_t) & 1 \\ -\lambda  & -i(u_x+u_t)
\end{array}\right],
\label{3.2}
\eeq
\beq
V=V(\lambda,x,t)=\left[\begin{array}{cc}
0 & -\frac{1}{\lambda}e^{iu} \\ e^{-iu} & 0.
\end{array}\right].
\label{3.3}
\eeq

As we mentioned above, the finite-gap ``spectral data'' consist of
\begin{enumerate}
\item A hyperelliptic Riemann surface $\Gamma$ with $2g+2$ branching points
$(0,\lambda_1,\ldots\lambda_{2g},\infty)$: $\mu^2=\lambda\prod\limits_{i=1}^{2g}(\lambda-\lambda_i)$
\item The divisor (a collection of points) $D=\gamma_1+\ldots+\gamma_g$ in $\Gamma$.
\end{enumerate}

In our text we always assume, that the spectral curve $\Gamma$ is {\bf generic}, i.e. all branching
points are  distinct.

Construction of the complex Sine-Gordon solutions is based on the following standard Lemma:

\begin{lemma}
For generic data $\Gamma$, $D$ there exists a unique two-component vector-function
$\Psi(\gamma,x,t)$ (the ``Baker-Akhiezer'' functions) such that
\begin{enumerate}
\item For fixed $(x,t)$ the function $\Psi(\gamma,x,t)$ is meromorphic
    in the variable
      $\gamma\in\Gamma$ outside the points $0$, $\infty$ and has at
      most 1-st order poles at the divisor points $\gamma_k$, $k=1,\ldots,g$.
\item  $\Psi(\gamma,x,t)$ has essential singularities at the points
      $0$, $\infty$ with the following asymptotic:
\beq
\Psi(\gamma,x,t)=\left(\begin{array}{c}
1+o(1) \\ i\sqrt{\lambda}+O(1)
\end{array} \right)e^{\frac{i\sqrt{\lambda}}{4}(x+t)} \ \ \mbox{as} \ \ \lambda\rightarrow\infty,
\label{3.4}
\eeq
\beq
\Psi(\gamma,x,t)=\left(\begin{array}{c}
\phi_1(x,t)+o(1) \\ i \sqrt{\lambda} \phi_2(x,t)+O(\lambda)
\end{array} \right)e^{-\frac{i}{4\sqrt{\lambda}}(x-t)} \ \ \mbox{as} \ \
\lambda\rightarrow 0,
\label{3.5}
\eeq
with some $\phi_1(x,t)$, $\phi_2(x,t)$.
\end{enumerate}
The Sine-Gordon potential $u(x,t)$ is defined by:
\beq
u(x,t)=i\ln \frac{\phi_2(x,t)}{\phi_1(x,t)}.
\label{3.6}
\eeq
\end{lemma}
Denote by $\lambda_k(x,t)$ the projections of the zeroes of the first component of $\Psi(\gamma,x,t)$ to the
$\lambda$-plane. Then
\beq
e^{iu(x,t)}=\prod\limits_{j=0}^{g}(-\lambda_j(x,t))\left/
{\sqrt{\prod\limits_{j=1}^{2g} E_j}}\right.
\label{3.7}
\eeq

{\bf Remark}. To be more precise, the formulas (\ref{3.1}-\ref{3.7}) define simultaneously a pair of
Sine-Gordon solutions $u_1(x,t)$, $u_2(x,t)$, depending on the choice of the branch $1/\sqrt(\lambda)$
near the point $\lambda=0$. They are connected by the following relation $u_2(x,t)=u_1(t,x)+\pi$. In the
real case it is possible to fix a canonical branch by making the analytical continuation along the real line.
This rule is unstable in the following sense: if we add a pair of complex conjugate branching points
which are very close to the positive half-line (or, equivalently, open a resonant point), it is a small
transformation in terms of the spectral data, but it exchanges $u_1$ with $u_2$..

The real Sine-Gordon solutions (by Cherednik's lemma they are automatically regular \cite{Cher})
correspond to the following data:
\begin{enumerate}
\item $\Gamma$ is real, i.e. the branching points of $\Gamma$ are either real, or form complex conjugate
pairs. Therefore we have an antiholomorphic involutions $\tau:(\lambda,\mu)\rightarrow (\bar\lambda,\bar\mu)$.
Denote the number of real finite branching points by $2k+1$.
\item All real branching points lie in the negative half-line $\lambda\le 0$. It is convenient to use  following
enumeration for the branching points different from $0$ and $\infty$: $0>\lambda_1>\lambda_2>\ldots>\lambda_{2k}$,
$\lambda_{2k+1}=\bar\lambda_{2k+2}$, \ldots, $\lambda_{2g-1}=\bar\lambda_{2g}$.
\item There exists a meromorphic differential $\Omega$ (Cherednik differential) with first order poles at $0$,
$\infty$, holomorphic on  $\Gamma\backslash\{0,\infty\}$ with the zeroes at the points $\gamma_1$,\ldots,$\gamma_g$,
$\tau\gamma_1$,\ldots,$\tau\gamma_g$ (or, equivalently the divisor $D$ satisfy the relation
$D+\tau D= 0+\infty -K$).
\end{enumerate}

As it was shown in \cite{Cher}, the variety of all real potentials corresponding to the given spectral curve
$\Gamma$ consists of $2^k$ connected components. A characterization of these components in terms of
the Abel tori was obtained  in \cite{Dubr-Nat} but this technique did not led to the the calculation
of topological charge through the inverse spectral data.

Our  calculation of the topological charge for the finite-gap Sine-Gordon solutions is based on the following
effective description of these components (see \cite{GN}, \cite{GN2}, \cite{GN3}):

Any meromorphic differential with first-order pole at $\infty$ can be written as:
\beq
\label{3.8}
\Omega=c\left(1-\frac{\lambda P_{g-1}(\lambda)}{R(\lambda)^{1/2}}\right)
\frac{d\lambda}{2\lambda},
\eeq
where $P_{g-1}(\lambda)$ is a polynomial of degree at most $g-1$. It is also natural to put $c=1$.
In case of the Cherednik differentials the set of zeroes is invariant with respect to $\tau$.  Therefore
all coefficients of the polynomial $P_{g-1}(\lambda)$ are real.

Assume, that we take an arbitrary real polynomial  $P_{g-1}(\lambda)$. Is it possible to construct a real
Sine-Gordon solution corresponding to it? The necessary and sufficient condition is the following:
{\bf the zeroes of $\Omega$ can be divided into two groups $\{\gamma_1,\ldots,\gamma_g\}$ and
$\{\gamma_{g+1},\ldots,\gamma_{2g}\}$ such, that $\tau\gamma_k=\gamma_{k+g}$, $k=1,\ldots,g$.} Equivalently,
a polynomial $P_{g-1}(\lambda)$ generates real SG solutions if and only if all real root of $\Omega$ have
even multiplicity. In generic situation (all roots form distinct complex conjugate pairs) each polynomial
$P_{g-1}(\lambda)$ generates $2^g$ different solutions. To choose one of them one has to say, which point to choose in
 each complex conjugate pair belonging to $D$ ( the second one belongs to $\tau D$ ).
In degenerate cases (i.e. if there are real roots) the number of choices is smaller. Al these solutions
associated with a given $P_{g-1}(\lambda)$ belong to the same real Abel torus.

{\bf Definition.} A polynomial $P_{g-1}(\lambda)$ (and the corresponding differential $\Omega$) are called
{\bf admissible} if all real roots of $\Omega$ have even multiplicity.

Admissible polynomials $P_{g-1}(\lambda)$ can be characterized in the following geometrical way:

Let us draw the graph of the functions
\beq
f_{\pm}(\lambda)=\pm\frac{\sqrt{R(\lambda)}}{\lambda},
\label{3.9}
\eeq
and  fill in the following domains by the black color:
\beq
\begin{array}{l}
\lambda<0, y^2 < \frac{R(\lambda)}{\lambda^2}, \\
\lambda>0, y^2 > \frac{R(\lambda)}{\lambda^2}.
\end{array}
\label{3.10}
\eeq

\begin{center}
\includegraphics[scale=0.3]{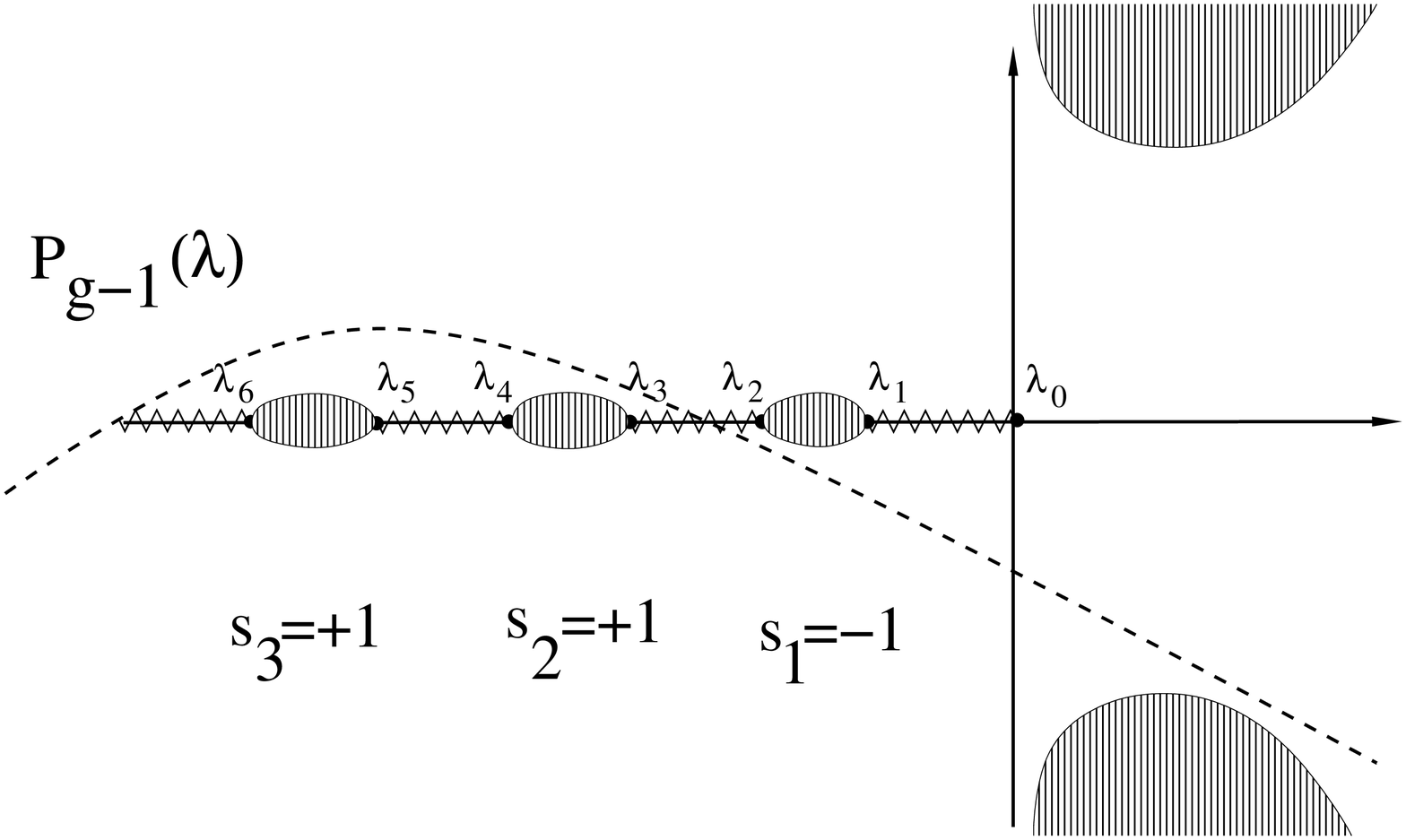}

Fig 1.
\end{center}

\begin{lemma}
\label{SG-L1} The polynomial $P_{g-1}(\lambda)$ is admissible if and only if the graph of
$P_{g-1}(\lambda)$ has no parts inside the black open domains.
\end{lemma}
If the  graph does not touch these domains, we have no real divisor points. Real divisor points correspond
to the case, when the graph touches one of these domains but does  not cross the boundary.

Each pair $\lambda_{2j-1},\lambda_{2j}$ is connected by a black ``island''. The graph of
admissible $P_{g-1}(\lambda)$ should go above or below this island, therefore at all intervals
$[\lambda_{2j},\lambda_{2j-1}]$, $j\le k$ $P_{g-1}(\lambda)\ne 0$. Let us associate with an admissible
polynomial $P_{g-1}(\lambda)$ a collection of numbers $s_j$, $j=1,\ldots,k$ by the following
rule: $s_j=1$ if the graph of $P_{g-1}(\lambda)$ is positive in the interval
$[\lambda_{2j},\lambda_{2j-1}]$ and $s_j=-1$ otherwise. Let us call the set $s_j$
{\bf topological type of the real solution}. We have exactly $2^k$ possible topological types.
Elementary analytic estimates (see \cite{GN2}) show, that all these components are non-empty. Each connected
component is a real Abel torus, and the $x$-dynamics defines a straight line in this torus.
To calculate the density of the topological charge it is sufficient to know the direction
of this line and the charges along the basic cycles. It follows from a simple analytic lemma:

\begin{lemma}
\label{SG-L0.1}
 Let $u(\vec X)$, $X\in{\mathbb R}^n$ be a smooth function in ${\mathbb R}^n$
 such, that $\exp(iu(\vec X))$ is single-valued on the torus
 ${\mathbb R}^n/{\mathbb Z}^n$.  Equivalently it means, that
 $\exp(iu(\vec X+\vec N))=\exp(iu(\vec X))$ for any integer vector $\vec N$, and
 \beq
 u(X^1,X^2,\ldots,X^k+1,\ldots,X^n)-u(X^1,X^2,\ldots,X^k,\ldots,X^n)=2\pi n_k.
 \label{3.11.1}
 \eeq
 The numbers $n_k$ are called {\bf the topological charges along the basic cycles
 $\frak A _k,k=1,\ldots,n$}.
 Denote by $u(x)$ restriction of $u(\vec X)$ to the strait line $\vec X=\vec X_0 +x\cdot\vec v$,
 $\vec v=(v^1,v^2,\ldots,v^n)$.
 Then the density of topological charge $\bar n=\lim\limits_{T\rightarrow\infty}
 [u(x+T)-u(x)]/2\pi T$ is well-defined; it does not depend on the point $\vec X_0$ and:
\beq
\bar n=\sum\limits_{k=1}^n n_k v^k.
\label{3.11}
\eeq
\end{lemma}

The calculation of the direction vector for the $x$-dynamics is absolutely standard (see, for example
\cite{Erc-Forest}). Denote by $\omega^l$ the canonical basis of holomorphic differentials on $\Gamma$:
\beq
\label{3.11.2} \omega^l=i\ \frac{\sum\limits_{j=0}^{g-1}
D^k_j\lambda^j}{\sqrt{R(\lambda)}}\ d\lambda, \ \ D^k_j\in{\Bbb R}
\eeq
Then for the components of the $x$-direction vector we have:
\begin{equation}
\label{3.11.3}
U_k=\frac{1}{2}\left(D^k_{g-1}+ {D^k_{0}}\left/{\sqrt{\prod\limits_{j=1}^{2g}E_j}}\right.\right).
\end{equation}

To obtain a simple expression for the basic charges it is critical to use
a proper basis of cycles in $\Gamma$. In \cite{GN}-\cite{GN3}) the authors used the following basis,
first suggested in \cite{DN}:

\begin{center}
\includegraphics[scale=0.3]{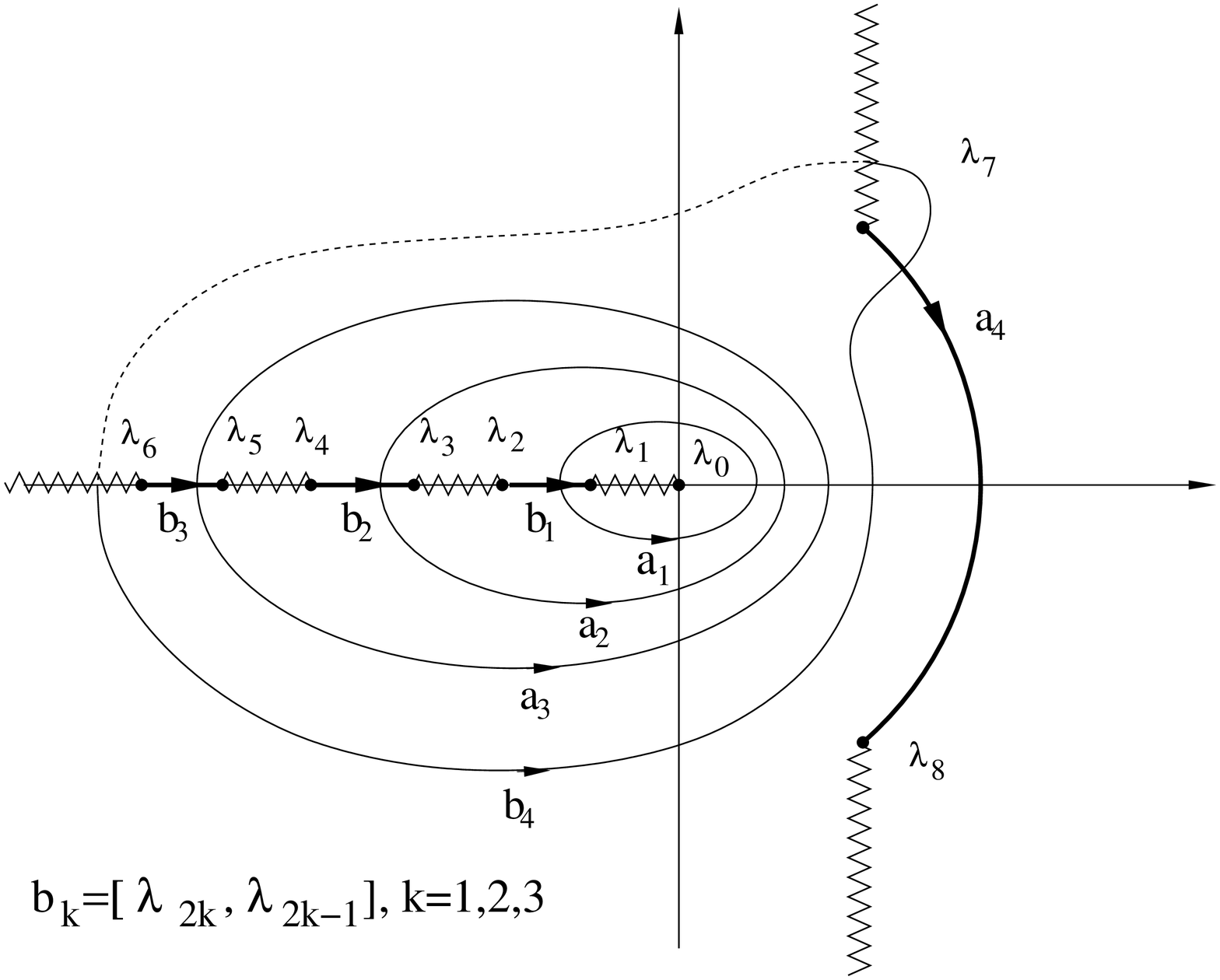}

Fig 2.
\end{center}
Here the cycles $a_j$, $j=1,\ldots,k$ are ovals on the upper sheet of $\Gamma$, containing inside
the points $\lambda_0=0$, $\lambda_1$, $\lambda_2$,\ldots$\lambda_{2j-1}$. The cycle $b_j$, $1\le j\le k$ lies over the interval
$[\lambda_{2j},\lambda_{2j-1}]$. The cycles $a_j$, $j=k+1,\ldots,g$ lie over pathes connecting the pairs
$\lambda_{2j-1}$ and $\lambda_{2j}$. We assume that these cycles do not intersect each other, and the cycles
 $a_j$, $j=k+1,\ldots,g$ do not intersect the negative semi-line. The cuts are shown by the zigzag lines.
The upper sheet contains the semi-line $\lambda>0,\mu>0$.

Consider a basic cycle ${\frak A}_j$ on the real component of Jacoby torus,
represented by the closed curve. The image of this cycle in $\Gamma$ under the inverse Abel map is
a closed oriented curve $C_j$, formed by the motion of the corresponding divisor points
(it may have several connected components). The motion of an individual divisor point
does not have to be periodic, after going along the cycle we may obtain a permutation of the divisor
points. The curve $C_j$  is homological to the cycle $a_j\in H_1(\Gamma,Z)$.
It follows from (\ref{3.7})  that the topological charge $n_j$ along the cycle ${\frak A}_k$ equals to
the winding number of the curve $C_j$ with respect to the point 0. Equivalently
\beq
\label{3.12}
n_j=\tilde C_j \circ {\mathbb R}_-,
\eeq
where $\circ$ denote the intersection number, $\tilde C_j $ denotes the projection of $C_j$ to the
$\lambda$-plane, ${\mathbb R}_-$ is negative semi-line with the standard orientation.

For each point of ${\frak A}_j$ the corresponding divisor $\gamma_1$,\ldots,$\gamma_{g}$ is admissible.
From the characterization of admissible divisors obtained above it is easy to show, that
the curve $C_j$ does not touch the closed segments on the real line
$[-\infty,\lambda_{2m}]$,\ldots,$[\lambda_3,\lambda_2]$, $[\lambda_1,0]$. Therefore any time the curve
$C_j$ crosses the negative semi-line, it intersects one of the basic cycles $b_j$, $j=1,\ldots,k$.

Unfortunately this information is not sufficient to calculate the basic charge, because the orientation
of the cycles $b_j$ coincides with the orientation of the negative semi-line at one sheet and they
are opposite at the other one. For example at the Fig. 3 we see two different realizations of the cycle
$a_1$ representng different topological types. $a_1$ is drawn at the upper sheet and $a'_1$ is drawn
at the lower one. We have $a_1\circ b_1 = a'_1\circ b_1=1$, but $\tilde a_1\circ {\mathbb R}_- =1$,
$\tilde a'_1\circ {\mathbb R}_- =-1$, therefore $n_1=1$ and $n_1=-1$ for these cycles respectively.

\begin{center}
\includegraphics[scale=0.5]{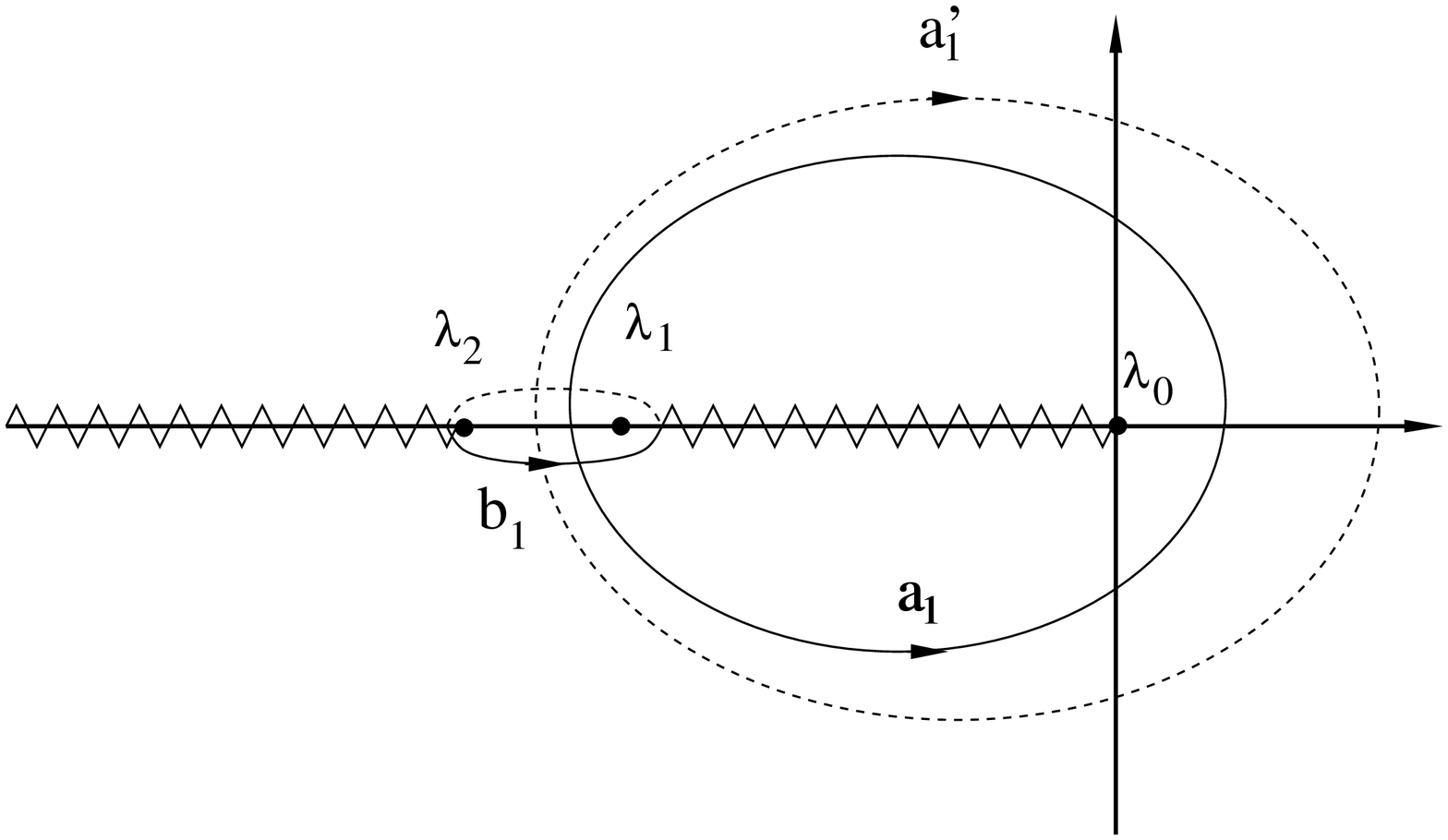}

Fig 3.
\end{center}

Fortunately, the topological type contains the information at which sheet the intersection takes
place. Namely we have:
\begin{lemma}
Assume, that the cycles $C_j$ intersects the negative semi-line at the interval $(\lambda_{2l},\lambda_{2l-1})$.
Then orientations of $b_l$ and ${\mathbb R}_-$ coincide in the intersection point if $(-1)^{l-1}s_l >0$ and
are opposite if $(-1)^{l-1}s_l < 0$.
\end{lemma}

Combining all these results we obtain the final formula:
\begin{theorem}
The density of the topological charge for a real Sine-Gordon solution is given by
\beq
\label{3.13}
\bar n=\sum\limits_{j=1}^{k} (-1)^{j-1} s_j U_j,
\eeq
where the vector $U_j$ is defined by (\ref{3.11.3}).
\end{theorem}

{\bf Acknowledgements}.

Petr Grinevich  was supported by the RFBR grant 04-01-00403a and by the grant NSh-4182.2006.1 of the Presidential
Council on Grants (Russia).

\end{document}